\documentclass[letterpaper,10pt]{article} 
\usepackage{opticameet3}
\newcommand\authormark[1]{\textsuperscript{#1}}
\usepackage{amsmath,amssymb}
\usepackage{physics}
\usepackage{bbold}
\usepackage[colorlinks=true,bookmarks=false,citecolor=blue,urlcolor=blue]{hyperref} 
\begin{document}
\title{Task-dependent semi-quantum secure communication in layered networks with OAM states of light }
\author{ Rajni Bala$^*$, Sooryansh Asthana$^{**}$, V. Ravishankar$^{\dagger}$}
\address{Department of Physics, IIT Delhi, Hauz Khas, New Delhi, 110016, Delhi, India}
\email{\authormark{*}Rajni.Bala@physics.iitd.ac.in,\authormark{**}sooryansh.asthana@physics.iitd.ac.in,\authormark{$\dagger$}vravi@physics.iitd.ac.in}
\begin{abstract}
Secure communication in layered networks having differently preferred participants has attracted a lot of research attention. Protocols for key distribution in a layered network have been recently proposed in [M. Pivoluska {\it et al.},
Phys. Rev. A 97, 032312] by employing asymmetrically entangled multiqudit states. Due to employment of asymmetrically entangled multiqudit states, the yield of these protocols is very low.  To address this issue, in this work, we have proposed semi-quantum secure communication protocols by employing separable states only which give a better yield and a higher key generation rate. As  illustrations, we present two representative protocols. The first protocol allows  sharing of two keys simultaneously in a  network  of two layers. The second protocol facilitates direct communication in one layer and key distribution in the other. The separable states, i.e.,  coherent pulses of orbital angular momentum required in the protocols are easily realisable with current technologies.  
\end{abstract}

\section{Introduction}
Recent times have witnessed an extensive development in secure quantum communication protocols (see\cite{pirandola2020advances} and references therein). More recently, quantum key distribution protocols for  multiple layers of a network have been proposed \cite{pivoluska2018layered}. In all these protocols, all the participants are supposed to have quantum capabilities, which might not be the case in realistic scenarios. This has led to proposal of the seminal semi-quantum key distribution protocol by Boyer {\it et al.} in 2007 \cite{Boyer07}. Since then, a large research interest is fostered in the proposals of semi-quantum secure communication protocols. In a more recent work \cite{bala2022layered}, we have presented protocols for simultaneous distribution of secure task-dependent information in multiple layers of a network. Multidimensional entangled states are used as a resource in these protocols. There has been certain experimental development in the generation of such states. However, to implement such protocols in large networks, the generation of requisite states is yet in its infancy \cite{malik2016multi}.

Such protocols can be realised with current technologies by employing separable states of light. In fact, by employing higher-dimensional separable states, key generation rate in each layer can also be increased.
As illustrations, we present two protocols for --(i) simultaneously distribution of keys in a network of two layers, (ii) secure transfer of a message in one layer and distribution of a key in the other layer. 

Higher-dimensional separable states of light are employed in the protocols which are readily available.  In fact, these protocols can be realised with coherent pulses of  orbital angular momentum (OAM) modes of light, thanks to many advancement in their generation and manipulation \cite{willner2021orbital}. Additionally, polarisation and OAM states of light as generated in \cite{PhysRevApplied.11.064058} can also be employed to implement these protocols. For this reason,  these protocols can be implemented with current state-of-the-art technology. These protocols do away with the need for employing multiqudit entangled OAM states, whose generation yield is very low \cite{malik2016multi} (of the order of several Hz). In addition, these protocols distribute information in a layered network, that too with higher key generation rate. So, they are natural candidate to be realised experimentally. 
\section{Semi-quantum key distribution in a network}
\label{SQKD}
Consider a network of three participants {\it viz.} Alice, Bob$_1$ and Bob$_2$ with two layers $L_1$ and $L_2$. Let Alice and Bob$_1$ belong to layer $L_1$ and all the three participants belong to layer $L_2$. Alice is the quantum participant, i.e., she can prepare any state and measure in any basis. However, Bob$_1$ and Bob$_2$ are classical participants, so their actions are restricted. They can prepare and measure a state only in the computational basis. 

In the protocol, Alice wants to distribute keys in both the layers with higher key generation rate. For this purpose, she prepares $9 \otimes 3$ dimensional separable states.\\
\noindent{\it Inputs:} To implement the task, Alice employs the following set of bases:
\begin{align}\label{eq:set}
 & S_1\equiv\big\{\ket{00},\ket{11},\ket{22},\ket{30},\ket{41},\ket{52},\ket{60},\ket{71},\ket{82}\big\},~~\nonumber\\
 & S_2\equiv\big\{\ket{0'0''},\ket{1'1''},\ket{2'2''},\ket{3'0''},\ket{4'1''},\ket{5'2''},\ket{6'0''},\ket{7'1''},\ket{8'2''}\big\}.
\end{align}
The states in $S_2$ are the Fourier transforms of the respective states in computational basis $S_1$. For this reason, the states in $S_1$ and $S_2$ are mutually unbiased. The steps of the protocol are as follows:
\begin{enumerate}
    \item Alice randomly prepares a state with an equal probability from  ${S}_1$ or ${S}_2$ and sends the first and the second subsystems to Bob$_1$ and Bob$_2$ respectively. She also notes down the quantum number of the first subsystem with herself.
    \item Each Bob, being a classical participant, exercises two choices with an equal probability. He either measures the received state in the computational basis  and sends the post-measurement state to Alice, or simply sends the received state back to Alice. 
    \item On receiving states from each Bob, Alice measures them in the same bases in which they were initially prepared. 
    \item This completes one round. After a sufficient number of such rounds, Alice reveals the rounds in which she had sent states from  $S_1$ and both the Bobs reveal the rounds in which they have performed measurements. 
    \item {\it Check for eavesdropping:}  Alice analyses data of the rounds in which none of the Bobs has performed a measurement. If there is no eavesdropping, Alice would get the same state. If eavesdropping is so ruled out, the data of those rounds in which Alice has sent a state from $S_1$ and both the Bobs have performed measurements constitute a key in the respective layer. 
    \end{enumerate}
    \subsection{Key generation rule} \label{key_rule}
To retrieve a key symbol in each layer, each participant writes her/ his outcomes in the base $3$ representation. Since Bob$_2$ has access to only a qutrit, his measurement results, represented by $b_2$, are already in the  base $3$ representation.
 Bob$_1$ has access to a nine-level system and Alice notes the quantum number of the state with her which she has sent to Bob$_1$. Both of them re-express it in base $3$ representation. 
Let $a,~b_1$ be the symbols with Alice, Bob$_1$ respectively, which can be written in the base $3$ form as:
\begin{equation}\label{outcomes}
    a=3a^{(1)}+a^{(0)},\quad b_1=3b_1^{(1)}+b_1^{(0)}.
\end{equation}
The symbols $a^{(1)},~b_1^{(1)}$ constitute a key in layer $L_1$ and the symbols $a^{(0)},~b_1^{(0)},~b_2$ constitute a key in layer $L_2$.\\
Since the key symbols are in the base $3$ representation and are generated with an equal probability,  the sifted key rate in each layer is ${\rm log}_23$ bits.
\section{Task-dependent layered semi-quantum secure communication (TLSQSC)}
\label{LSQSDC+KD}
In the previous section, we have shown that the task of simultaneous key distribution in a layered network, that too with a higher key generation rate, can be implemented with higher-dimensional separable states of light. In this section, we aim to show that the same set of states (equation (\ref{eq:set})) also allows to realise task-dependent communication in the same network. The requirement is to establish a secure but direct communication in layer $L_1$ and distribution of a key in layer $L_2$ is to be realised.
The steps of the protocols are the same as those of SQKD, albeit with a small modification which is described below.
\begin{enumerate}
\item The steps $(1-5)$ are same as those of SQKD.
\item After checking for the absence of eavesdropper, Alice shares a sequence of symbols $m \equiv M\oplus_3 b_1^{(1)}$ on an authenticated classical channel with Bob$_1$ to transfer her message $M$. The symbol $b_1^{(1)}$ is the same as that of equation (\ref{outcomes}).  The symbol $\oplus_3$ represents addition modulo $3$. 
\item Since Bob knows $b_1^{(1)}$, he retrieves the message $M$ by applying decoding $M=m\oplus_3 \Tilde{b_1}^{(1)}$, where $\Tilde{b_1}^{(1)}$ is inverse of $b_1^{(1)}$ under addition modulo $3$.

    \end{enumerate}
In this manner, Alice transmits her message securely in layer $L_1$ and distributes a key in layer $L_2$.

\noindent{\bf Confidentiality in the communication:} The communication between Alice and Bob$_1$ is completely confidential, shielding it from Bob$_2$. This is because, Bob$_2$ is denied access to the outcomes of Bob$_1$'s measurements. 
\section{Scope for experimental implementation with orbital angular momentum modes of light}
OAM modes of light provide a platform to physically realise higher-dimensional states. There has been a great advancement in  generation and measurement of OAM modes of light since its discovery \cite{willner2021orbital}. 

\noindent{\it Generation of states:} To generate the states for implementing the proposed protocols, it is required to generate states from the bases $S_1$ and $S_2$ (equation (\ref{eq:set})).  Since ${S}_1$ is the computational basis, the states belonging to it can be identified with the Laguerre-Gauss modes of light. The basis ${S}_2$ is mutually unbiased with respect to the basis ${S}_1$. So, the states belonging to the bases ${S}_2$  involve coherent superposition of nine and three Laguerre Gauss modes respectively which can be realised with spatial light modulators. The optical setup for generating coherent superposition of four Laguerre Gauss modes with equal weights has already been given in \cite{PhysRevApplied.11.064058}.\\
{\it Measurement of states:} Each Bob can perform the measurement in the Laguerre Gauss basis by performing a log-polar transformation.  A measurement can also be performed by employing spatial light modulator (SLM) and detectors or with the techniques  presented in \cite{mirhosseini2013efficient}.\\

\section{Conclusion}
In summary, we have shown that by employing separable multi-dimensional states, the so-called next generation protocols for distribution of keys in a layered network can be realised with current technologies. In addition, employment of higher-dimensional states provides an increase in the key generation rate in different layers of network, making them more efficient. Our protocols may be implemented with faint coherent pulses as well, thereby alleviating the need for spontaneous parametric down conversion, whose yield is very low.


\end{document}